\documentclass{mn2e}
\usepackage{psfig,times}

\title[The properties of galaxies at $z>5$]{The properties of galaxies at $z>5$}

\author[M.N. Bremer et~al.]{
M.N. Bremer$^1$, M.D. Lehnert$^2$, I. Waddington$^1$, M.J. Hardcastle$^1$, P.J. Boyce$^1$, S. Phillipps$^1$\\
$^1$Department of Physics, Bristol University,H.H. Wills Laboratory,
Tyndall Avenue, Bristol BS8 1TL, U.K.\\
$^2$Max-Planck-Institut f\"ur extraterrestrische Physik,
Giessenbachstra\ss e, 85748 Garching bei M\"unchen, Germany }

\begin{document}
\def\Msun{\hbox{$\rm\thinspace M_{\odot}$}}
\def\Msunperyr{\hbox{$\Msun\yr^{-1}\,$}}
\def\Mdot{\hbox{$\dot {M}$}} 
\maketitle

\begin{abstract}

In a recent paper Lehnert \& Bremer have photometrically selected
a sample of galaxies at $z>4.8$ from a single VLT/FORS2 pointing and
spectroscopically confirmed half of them to be at $4.8<z<5.8$. To study
the properties of such galaxies further, we have photometrically selected
a similar sample ($V_{AB}>28, i_{AB}<26.3, i_{AB}-z_{AB}>0$) from the HST
ACS images of the {\it Chandra} Deep Field South. This selection results
in a sample of 44 sources from $\sim 150$ arcmin$^2$. We find that such
galaxies are often barely resolved in the ACS images, having half-light
radii of 0.1-0.3 arcsec ($<2$kpc). They show no difference in spatial
clustering from sources selected by $i_{AB}<26.3, i_{AB}-z_{AB}>0$, which are
generally galaxies of lower redshift. However, their distribution over
the field is not uniform and their surface density varies considerably
over areas comparable to a single 8m or HST pointing. A reliable
determination of the surface and volume densities of such galaxies
requires a sky area considerably larger than the current ACS imaging of
this field.  No individual $z>5$ candidate was detected to a 3-$\sigma$
limit of $6\times 10^{-17}$ erg s$^{-1}$ cm$^{-2}$ at 0.5-5 keV by {\it
Chandra} (a limiting luminosity of below $2 \times 10^{43}$ erg s$^{-1}$
at $z\sim5.3$). By summing over all positions, we find that the mean
source must be undetected at a level at least a factor 4 times fainter
than this.  This rules out anything other than a weak AGN contribution
to the emission from these objects and thus luminous AGN made little
contribution to the final stages of re-ionization of the Universe.

\end{abstract}


\vspace{-7mm}
\section{Introduction}
\vspace{-2mm}

With the advent of high throughput cameras on 8m-class telescopes and
the Advanced Camera for Surveys (ACS) on the Hubble Space Telescope it
is now possible to detect and study galaxies at, or close to the epoch
of reionization, about 1 billion years after the Big Bang. Recently,
many such galaxies have been selected photometrically ( {\it e.g.}
Stanway et~al., 2003, Bouwens et~al., 2003, Iwata et~al., 2003) and some
have been confirmed to be at high redshift by spectroscopy (Lehnert \&
Bremer, 2003; Bunker et~al., 2003; Cuby et~al., 2003)

In a recent paper, we used perhaps the simplest technique to select
and confirm redshifts for a high redshift galaxy sample drawn from a
single 40 arcmin$^2$ field imaged by the VLT (Lehnert \& Bremer
2003). The opacity of neutral hydrogen along the line-of-sight to the
galaxies means that the light shortward of 1216 \AA ~ in their
rest-frame is strongly absorbed, leading to a sharp drop in their
spectra at the corresponding redshifted wavelength. By selecting
objects detected in the $I-$band to $I_{AB}=26.3$, but undetected at
$R_{AB}>27.8$ ($i.e.$ R-band ``dropouts''), we identified 13 candidate
$z>4.8$ galaxies. Spectroscopy of 12 of these confirmed that 6 were at
$4.8<z<5.8$, as they possessed a Lyman $\alpha$ emission line at the
same wavelength as a break in their spectra. The spectra were
consistent with the objects being strongly starforming
galaxies. Similar techniques have been used by other groups to
identify other unobscured star forming galaxies at even higher
redshifts (Stanway et~al. 2003, Bouwens et~al., 2003).  From the study
of galaxies at z$>$5, Lehnert \& Bremer (2003) concluded that
relatively luminous galaxies ($M_{AB}(1700\AA)>-21$) have insufficient
numbers and ionizing luminosity to keep the universe at z$>$5 ionized
and, since no broad line objects were detected, that QSOs contribute
very little to the overall ionization.  This last conclusion is
supported by Barger et al. (2003) who found that the number of
luminous X-ray-selected QSOs in the Chandra Deep Field North is
insufficient to contribute greatly to the overall ionization of the
IGM at z$\approx$5.5.

However, our original study was limited by the size of the field imaged
by the VLT (40 arcmin$^2$), the ground-based seeing (0.8-0.9 arcsec)
of the imaging and spectroscopy, and the lack of multi-wavelength data
for this field. Of particular concern was the AGN fraction in faint
galaxies at these redshifts and the possible contribution of ``hidden
AGN'' to the overall ionization budget of the high redshift universe.
Classes of AGN exist that have strong X-ray emission with the spectral
characteristics of AGN but with only subtle or no signs of AGN activity
in the rest-frame UV or optical (e.g., Giacconi et al. 2001). However,
by performing a comparable selection on a larger field imaged by HST and
other telescopes over a range in wavelengths, we can determine scale
sizes for such sources, determine if any are AGN, and examine their
spatial distribution and clustering. All these properties are required
for comparisons between the observations and models of the evolution
of the earliest galaxies. To this end in this paper we discuss the
properties of candidate $z>5$ galaxies in the Chandra Deep Field South
(CDF-S). In this paper, unless noted otherwise all magnitudes are on
the AB scale, and the cosmology used is H$_0=70$kms$^{-1}$ Mpc$^{-1}$,
$\Omega_{\Lambda}=0.7$ and $\Omega_M=0.3$.

\section{Sample Selection}

In this paper we use data from the Advanced Camera for Surveys
(ACS, Ford et al. 2003) on HST, released as part of the public Great
Observatories Origins Deep Survey (GOODS - Dickinson \& Giavalisco
2002) programme. Specifically, we utilise data from the v0.5 release
of GOODS data from the Chandra Deep Field South (CDF-S). The data in
this release comprises that from the first 3 epochs of the CDF-S survey.
At each epoch, a given field is imaged in three of the four bands used in
the survey: $V$ (F606W), $i$ (F775W) and $z$ (F850LP).  Observations in
the $B-$band (F435W) were all taken during epoch 1. The field is mosaiced
into a pattern of 15 (epochs 1 and 3) or 16 (epoch 2) 'tiles'. Exposure
times at each epoch in each tile are 0.5, 0.5 and 1 orbits in the $V$,
$i$ and $z$ bands respectively.  Exposure time in the $B-$band (epoch 1)
was 3 orbits per tile. At each epoch, the observations of each tile
are broken into 2, 2 and 4 frames, in $V$, $i$ and $z$ respectively,
to allow for the removal of cosmic rays and other defects. The B-band
observations are broken into 6 frames. The released data set includes
pipeline-processed, cosmic-ray cleaned, drizzled (Fruchter \& Hook 2002)
and co-added frames of each tile for each epoch for each of the 4 bands.

The tiles from epochs 1 and 3 are based on the same mosaic pattern
covering the total survey area, although the epoch 3 frames have a PA 90
degrees offset from those of epoch 1. The tiles of epoch 2 are based on
a different mosaic pattern (of 16 tiles). Hence, we decided to create
co-added $V$, $i$ and $z$ tiles from each of the sets of 15 tiles in
epochs 1 and 3. The $z$ tile from epoch 1 was used as a reference tile
in each case. The $z$ tile from epoch 3, and the $V$ and $i$ tiles
from epochs 1 and 3, and the $B$ tiles from epoch 1 were aligned to the
equivalent epoch 1 $z$ tile using the wcsalign task in KAPPA.  Each pair
of co-aligned $z$, $i$ and $V$ frames were then co-added using the makemos
task in CDDPACK. For each co-added image, all areas which had blank values
in either input image were set to blank in the final co-added image. This
included the gap between the two ACS CCDs on the $i$, $z$ and $V$ frames.

As we are searching specifically for objects which are securely
detected in the redder bands, a catalogue of sources was constructed
based on the 15 co-added $z$ tiles. The detection and photometry of
objects in the co-added $z-$band tiles were conducted automatically
using the SExtractor software package (Bertin \& Arnouts 1996) as
implemented within the Starlink GAIA image display and analysis
package. For object identification we demanded at least 4 contiguous
pixels above a threshold of 1.5$\sigma$ per pixel. We measured Kron
(1978) magnitudes taken within an aperture radius of 2.5r$_{\rm
kron}$. These magnitudes have not been corrected for the extremely low
Galactic extinction of $A_V=0.024$.  The positions and apertures
defined by the SExtractor source detection algorithm as run on the
co-added $z-$band tiles were then used on the equivalent $i$, $V$ and
$B$ frames. In this way we measured the apparent $i$, $V$ and $B$
magnitudes for those objects detected on the $z$ tiles inside
identical apertures. We used the zero-points for AB magnitudes
determined by the GOODS team: mag$_{\rm AB}$ = zeropoint - 2.5
log(Count rate/s$^{-1}$) where the zeropoints were 25.662, 26.505,
25.656, 24.916 for the $B$, $V$, $i$ and $z$ bands respectively.

Having created and verified the catalogue, we then selected candidate
high redshift galaxies. Lehnert \& Bremer (2003) selected objects
with $25<I_{AB}<26.3$, and $R_{AB}-I_{AB}>1.5$, with all of the
final selection having $R_{AB}>27.8$. They noted that at $z>4.8$, the
predicted $I-z$ colours of galaxies became increasingly red, from $\sim
0$ at $z<4.8$ to $\sim 1.3$ at $z>5.8$. The observed colours of the
spectroscopically-confirmed high redshift galaxies roughly agreed with
this. Stanway et~al. (2003) have used a selection of $i_{AB}-z_{AB}>1.5$
to select $z\sim 6 $ candidates from the ACS data of the CDF-S (see
their Figure 3).

\begin{table}
\begin{tabular}{cllcccc}
  Number &    RA   &  Dec & $i$ & $V-i$ & $i-z$ & R$_h$ \\
    1  & 03 32 08.68  &-27 43 14.6 & 25.9 & $>$2.1 & 0.7    & 0.35\\
    2  & 03 32 11.26  &-27 46 30.4 & 26.3 & $>$1.7 & 0.3    & 0.18\\
    3  & 03 32 15.40  &-27 52 26.3 & 26.1 & $>$1.9 & 0.1    & 0.13\\
    4  & 03 32 16.16  &-27 46 41.5 & 26.2 & $>$1.8 & 0.1    & 0.20\\
    5  & 03 32 16.91  &-27 48 08.2 & 26.3 & $>$1.7 & 0.2    & 0.27\\
    6  & 03 32 17.47  &-27 50 03.1 & 26.1 & $>$1.9 & 0.8    & 0.27\\
    7  & 03 32 17.80  &-27 50 52.6 & 25.7 & $>$2.3 & 0.7    & 0.30\\
    8  & 03 32 18.18  &-27 47 46.5 & 25.3 & $>$2.7 & 1.5    & 0.10\\
    9  & 03 32 19.22  &-27 45 45.5 & 24.9 & $>$3.1 & 1.3    & 0.10\\
   10  & 03 32 19.74  &-27 41 40.1 & 26.2 & $>$1.8 & 0.6    & 0.27\\
   11  & 03 32 20.94  &-27 53 39.0 & 25.3 & $>$2.7 & 0.3    & 0.24\\
   12  & 03 32 21.29  &-27 40 51.3 & 25.9 & $>$2.1 & 0.3    & 0.20\\
   13  & 03 32 22.46  &-27 50 47.1 & 26.3 & $>$1.7 & 1.7    & 0.10\\
   14  & 03 32 22.69  &-27 51 54.1 & 26.2 & $>$1.8 & 0.2    & 0.32\\
   15  & 03 32 23.98  &-27 41 08.0 & 26.1 & $>$1.9 & 0.7    & 0.13\\
   16  & 03 32 27.56  &-27 56 26.3 & 25.3 & $>$2.7 & 0.1    & 0.30\\
   17  & 03 32 28.87  &-27 41 32.6 & 26.1 & $>$1.9 & 0.2    & 0.45\\
   18  & 03 32 30.57  &-27 42 43.5 & 25.8 & $>$2.2 & 0.9    & 0.20\\
   19  & 03 32 31.07  &-27 51 17.7 & 25.8 & $>$2.2 & 0.1    & 0.20\\
   20  & 03 32 31.09  &-27 42 26.8 & 26.1 & $>$1.9 & 0.9    & 0.35\\
   21  & 03 32 31.74  &-27 54 13.9 & 26.2 & $>$1.8 & 0.0    & 0.17\\
   22  & 03 32 33.47  &-27 50 29.9 & 26.2 & $>$1.8 & 0.4    & 0.15\\
   23  & 03 32 35.10  &-27 51 25.2 & 26.1 & $>$1.9 & 0.1    & 0.20\\
   24  & 03 32 36.23  &-27 43 15.3 & 26.0 & $>$2.0 & 0.4    & 0.37\\
   25  & 03 32 37.18  &-27 51 10.9 & 26.0 & $>$2.0 & 0.1    & 0.30\\
   26  & 03 32 37.25  &-27 42 02.7 & 26.0 & $>$2.0 & 0.3    & 0.25\\
   27  & 03 32 37.90  &-27 53 30.9 & 26.2 & $>$1.8 & 0.8    & 0.18\\
   28  & 03 32 40.40  &-27 51 42.4 & 26.0 & $>$2.0 & 1.0    & 0.25\\
   29  & 03 32 41.22  &-27 43 10.0 & 26.2 & $>$1.8 & 0.9    & 0.45\\
   30  & 03 32 42.33  &-27 42 04.1 & 25.6 & $>$2.4 & 0.6    & 0.27\\
   31  & 03 32 42.60  &-27 54 28.8 & 26.3 & $>$1.7 & 0.4    & 0.32\\
   32  & 03 32 42.62  &-27 54 28.9 & 26.3 & $>$1.7 & 0.7    & 0.30\\
   33  & 03 32 42.65  &-27 49 39.0 & 26.0 & $>$2.0 & 0.3    & 0.20\\
   34  & 03 32 43.29  &-27 43 10.7 & 26.1 & $>$1.9 & 0.1    & 0.35\\
   35  & 03 32 45.51  &-27 52 50.3 & 26.1 & $>$1.9 & 0.3    & 0.35\\
   36  & 03 32 47.64  &-27 51 05.0 & 26.3 & $>$1.7 & 0.7    & 0.10\\
   37  & 03 32 48.13  &-27 48 17.7 & 25.8 & $>$2.2 & 0.4    & 0.27\\
   38  & 03 32 49.14  &-27 50 22.5 & 26.2 & $>$1.8 & 0.7    & 0.10\\
   39  & 03 32 50.66  &-27 46 30.5 & 26.1 & $>$1.9 & 0.0    & 0.15\\
   40  & 03 32 52.77  &-27 51 25.7 & 25.4 & $>$2.6 & 1.2    & 0.13\\
   41  & 03 32 52.87  &-27 54 55.9 & 26.2 & $>$1.8 & 0.2    & 0.45\\
   42  & 03 32 53.95  &-27 54 52.3 & 26.2 & $>$1.8 & 0.0    & 0.18\\
   43  & 03 32 58.65  &-27 52 43.7 & 26.1 & $>$1.9 & 0.0    & 0.15\\
   44  & 03 32 59.70  &-27 52 02.6 & 26.1 & $>$1.9 & 0.2    & 0.20\\

\end{tabular}
\caption{ Coordinates are J2000. All magnitudes are in AB. Errors on
magnitudes are typically 0.1 in $i$.  $V-i$ colours determined
assuming $V>28$. Errors on $i-z$ are typically 0.15. R$_h$ is the
half-light radius measured in arcseconds. Unresolved sources have
$R_{h}=0.1$, although given the uncertainties on determining the
half-light radii of objects at $i\sim26$, the uncertainty on this
measurement should be assumed to be approximately 1 pixel or 0.05
arcsec.}
\end{table}

\begin{figure*}
\centerline{\psfig{file=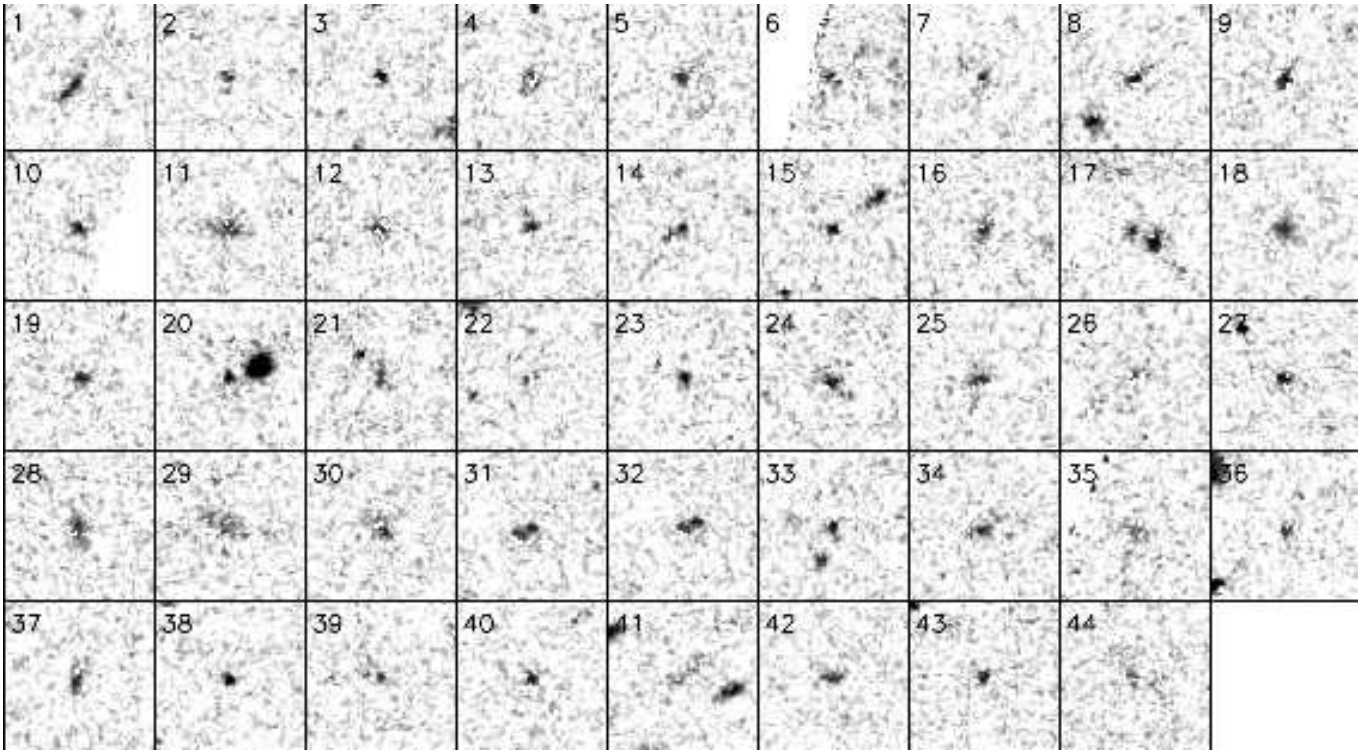,width=18cm}}
\caption{3$\times$3 arcsec $Viz$-composite images of all 44 objects in
the sample. In each case the selected object is at the centre of the
image and north is at the top and east is to the left.}

\end{figure*}

However, the ACS data were obtained in $B,V,i$ and $z-$bands. Consequently
we had to select galaxies that dropped out in $V$ rather than in $R$. This
will potentially select galaxies at $z>4.5$, rather than at $z>4.8$ as
did the $R-$ band dropout selection of our previous work. We therefore
also required that $i_{AB}-z_{AB}>0$ to minimise the number of lower
redshift dropouts. We also required that there was no sign of a detection
in $V$, placing a limit of $V_{AB}>28$ in the aperture matched to the
$z-$band. Objects brighter than this could always be detected in the
$V-$band frame. The $V-$band number counts show  that $V>28$ marked a
transition from true detection to non-detection in our catalogue. Thus
our final selection was $V_{AB}>28$, $i_{AB}<26.3$ (to be consistent
with our previous work) and $i_{AB}-z_{AB}>0$ (in order to reject lower
redshift dropouts).

Based upon the above magnitude and colour cuts, we selected objects
from our catalogue. We examined each of them in turn in order to
exclude obvious problem or bogus sources, such as parts of larger
sources or spurious sources at the edges of CCD frames where extreme
colours were caused by missing counts in one band off the edge of the
frame. The final selection consisted of 44 objects spread over the
$\sim 150 $ arcmin$^2$ of the ACS image of the CDF-S. Images of these
sources are shown in Figure 1 and their properties are given in Table
1. Note that in principle, with our selection criteria, we could have
included two of Stanway et~al's $z \sim 6$ objects; one is indeed
detected (their object 4, our object 8), the other (their object 3)
just misses our magnitude cut (due to our re-measurement of the
$i-$band photometry).

What contamination do we expect in our catalogue from sources other
than galaxies at $z>5$? At magnitude levels above $i_{AB}\sim 24-25$
we expect some cool stars with red $V-I$ and $I-z$ colours ({\it e.g.}
see Fig 2 in Iwata et~al., 2003). Based upon the results in Lehnert \&
Bremer (2003) we can expect up to $\sim 10$ such stars brighter than
$i_{AB}=25$ (both the CDF-S and the field in Lehnert \& Bremer are at high
Galactic latitudes), although many of these may be rejected by our strong
$V-$band limit. The work of Stanway et~al. (2003) on the CDF-S indicate
there are only one or two sub-stellar objects in the field at fainter
magnitudes. Given our previous results, and those of Stanway et~al,
our colour selection will also detect several Extremely Red Objects
(EROs) in the field, either old ellipticals or reddened galaxies at
around $z\sim 1$ ({\it c.f.} Cimatti et~al 2002). These objects will
appear resolved on scales of $>0.3$ arcsec.

Some publically-available ground-based data in $R$ and $J$ and $Ks$
are available from the VLT for part of this field. We can use these
data to attempt an estimation of the likely contamination from lower
redshift sources, even though the data are not ideal for this. The
$R$-band data is shallower than that used by Lehnert \& Bremer (which
went to $R_{AB}=27.8$) to select $R-$ band dropouts. For the CDF-S,
$3-\sigma$ detection limits in 2 arcsecond diameter apertures vary
between $R_{AB}=27.4$ and $26.7$ depending upon the particular
image. Nevertheless the data has some utility in ruling out $V-$band
dropouts at below $z=4.8$ and lower redshift red galaxies. About 15
per cent of the candidate distant galaxies are detected (4 were
detected out of 27 imaged in R-band), including objects 24, 34 and 42
at $R=26.5 \pm 0.2$, $R=26.2 \pm 0.2$ and $R=26.4\pm 0.3$ respectively
which are among the galaxies with the largest half-light radii.  This
supports our contention that the galaxies with the largest half-light
radii are likely to be lower redshift galaxies.

\begin{figure}
\psfig{file=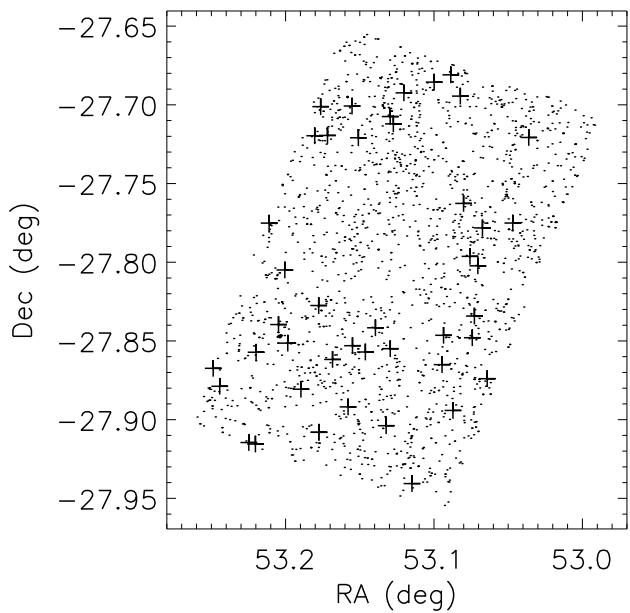,width=9cm}
\caption{Spatial distribution of our $z>5$ candidates over the
field (crosses). Underlying points are 2000 objects chosen to have
$i_{AB}<26.3$ and $i_{AB}-z_{AB}>0$. Note the $\sim 25$ arcmin$^2$
``hole'' in the distribution of candidates (centred at approximately
RA=53.15, Dec=-27.77), not reflected in the distribution of the non
dropout sources.}

\end{figure}

The $J$ and $Ks$ images also covered only part of the ACS area. The
data is useful to detect intrinsically red objects, as opposed to
those with breaks between $V$ and $i$.  Ten of our candidates were
covered by this data. Six were reliably detected in both $J$ and $Ks$
to $2-\sigma$ detection limits of $J_{Vega}=24.5$ and $Ks_{Vega}=23.5$
in 1 arcsec apertures.  These all had $z_{AB}-J_{Vega}$ colours $>1.6$
and $J-Ks$ colours ranging from 1.3 to 2.6. These are the colours of
intrinsically red objects and redder than expected for a galaxy formed
within a few hundred Myrs of $z=5.3$. In order to determine the
near-IR colours we might expect from an unreddened galaxy at $z\sim
5.3$, we ran a series of PEGASE models (Fioc \& Rocca-Volmerange 1997)
of ellipticals which started forming stars between $z=6.5$ and
$z=20$. $z_{AB}-J_{Vega}$ colours varied between $\sim 0.8$ and
$\sim1.2$ and $J-Ks_{Vega}$ colours between $\sim 0.9$ and
$\sim1.6$. These are similar colours to those of Irregular galaxies
redshifted to $z>5$, see {\it e.g.} Figure 4 of Stanway
et~al. (2003). Thus the six red objects appear redder than these
models of high redshift galaxies with unobscured ongoing star
formation.

The other six objects covered by this imaging were either undetected
or had colours consistent with being unreddened $z>5$ galaxies. What
implication does this have for the contamination of our $z>5$
candidate sample by low redshift red objects? The $J$ and $Ks$ data
only cover part of the ACS field, we need to use a proxy for red
near-IR colours in order to estimate the contamination. Five of the
six objects with $z_{AB}-J_{Vega}$ and $J-Ks_{Vega}$ colours redder
than the model galaxies also had $i-z \geq 0.8$; we can simply assume
that all objects in our sample with such a colour are potentially
intrinsically red or reddened objects at some arbitrary
redshift. There are nine such objects in our sample. Thus under this
assumption, we might expect up to about 20 per cent contamination of
our $z>5$ candidate sample by lower redshift red objects. For now we
choose not to remove these objects, as these could potentially be
obscured AGN at $z>5$ (the reddened nucleus dominating at the longer
wavelengths). As we will consider later the number of potentially
obscured AGN in our sample, by cross-correlation with the {\it
Chandra} data for the field, we will retain the objects in the full
sample.

Thus, between the $R-$band detected sources and those with red near-IR
colours we have a contamination rate of no more than 35 per cent
(about 15 sources). Some of the reddest objects could still be at
$z>5$ and there could be overlap of this sample with the $R-$band
detected sample (we cannot tell with the current data as the two
ground-based fields do not overlap effectively). Thus contamination by
lower redshift objects could be somewhat less than this. For a smaller
sample of higher redshift candidates, Stanway et~al. estimated a
contamination rate of about 25 per cent.

\section{Surface density and distribution of the objects}

The density of candidates with these colours and magnitudes, 1 per 3
arcmin$^2$, is similar to that found by Lehnert \& Bremer for their
sample of 13 candidates and is approximately twice as high as for
their spectroscopically confirmed $z>4.8$ galaxies. It is not clear
how to compare these surface densities, given that both samples could
be contaminated by lower redshift objects. As was argued in Lehnert \&
Bremer (2003), the high success rate of spectroscopic confirmation
would seem to indicate that most of the unconfirmed high redshift
candidates are also at $z>4.8$ as only about 50 per cent of lower
redshift dropout samples appear to have strong Lyman $\alpha$
emission.  In addition, Lehnert \& Bremer were able to
spectroscopically confirm that many of the color-selected sources at
the bright end of their sample (I$\la$24.5) were not high redshift
galaxies or QSOs.

A more likely problem is contamination in our CDF-S source list. We cannot
fully exclude EROs on the basis of our photometric cuts and there could be
slightly reddened lower redshift dropouts in our sample.  In any event,
the completeness of the ACS sample studied here will be higher as the ACS
data are deeper in $i$ and $z$ and have much higher spatial resolution
than the VLT data (0.08 arcsec as opposed to 0.8 arcsec).

Rather than attempt to determine a comoving volume density for these
sources, our purpose here is to note merely how the candidates are
distributed across the field. Figure 2 shows that they appear to cluster,
or at least show a wide range in their surface density.  One area of
$\sim 25$ arcmin$^2$ shows no detected sources. We determined that the
total catalogue of all objects showed no such ''hole'' or under-density in
sources in the same area. We also checked the photometry and number counts
in each band in this area against the rest of the catalogue to ensure
the lack of dropouts was not caused by problems in the photometry. No
differences were found. We compared the nearest-neighbour statistics of
the sources to random selections of 44 sources taken from our catalogue
using the cuts $i_{AB}<26.3, i_{AB}-z_{AB}>0$, but with no $V$-band
cut. This selection would tend to select galaxies at lower redshifts than
the dropout sample. Using a K-S test we found no significant difference
between the distributions of nearest neighbours for the high redshift
sample and the randomly-selected samples. Due to the low surface density
of the randomly-selected samples, they sometimes showed similar-sized
areas devoid of sources to that seen in the distribution of our high
redshift candidates.

Given the current data, there appears no obvious difference in
the clustering characteristics of our $z>5$ sample and other faint
sources. However, the fact that there is such a wide variation in the
surface density on scales of about 25 arcmin$^2$ indicates that a proper
treatment of the clustering properties and densities of these sources
requires imaging over an area considerably larger than that covered by the
ACS CDF-S data. Given that 25 arcmin$^2$ is comparable in size to a single
pointing of 8m telescope cameras such as FORS and GMOS, and is larger than
a single pointing of the ACS, caution must be exercised over interpreting
clustering properties and source densities derived from single pointings.

The projected proper area of the field is about 20 Mpc$^2$ at $z \simeq
5$. If we very crudely estimate that we probe of order 0.5 in redshift
around $z=5$ (in fact the effective volume will be less than this as
the completeness at faint levels is always less than 100 per cent, {\it
e.g.} see Lehnert \& Bremer), the comoving volume covered by the field
is roughly $10^5$ Mpc$^3$. Given the ``hole'' in the distribution of
candidate $z>5$ sources is of order 10-20 per cent of the imaged area,
this corresponds to a ``void'' with a size of order $\sim$ 10$^4$
Mpc$^3$. Thus it is not surprising that significant ``holes'' in the
distribution exist, as they correspond to relatively small comoving
volumes.

\section{The Morphologies of the Sources}

About half of the sources in Table 1 and Figure 1 are convincingly
resolved, but with half light radii of typically no more than 0.2-0.3
arcsec ($<2$ kpc). Several sources are unresolved but at least two are
resolved on rather larger scales of order 0.4-0.5 arcsec (17 and
41). These latter sources are most likely red $z\sim 1$ galaxies (41,
for example, has a secure R-band detection as noted in \S 2).  Several
sources appear as a double source or with a nearby lower surface
brightness component, though usually only a single component makes our
magnitude cut. An exception to this is the pair with catalogue numbers
31 and 32. Such sources appear similar to the pair of $z=5.3$ galaxies
discovered in the HDF-N by Spinrad et~al (1998). Overall, the sizes of
the sources are comparable to those found by Stanway et~al. for their
$i-$band dropouts, showing clearly that galaxies at $z>5$ have
scale-lengths of only $<$0.5 to 2 kpc (assuming a redshift range of
4.8-5.8 and our adopted cosmology).  Lowenthal et~al. (1997) noted
that $z=3$ dropouts in the HDF-N had typical half light radii of
0.2-0.4 arcsec (2-3 kpc in our assumed cosmology). Results from Roche
et~al., (1998) indicate that typical $z=4$ galaxies have half light
radii smaller than those at lower redshift. Their $z=4$ half-light
radii are comparable to ours at $z>5$ (when transformed to our
cosmology), so our results appear to support a decrease in galaxy
scale-lengths with increasing redshift. It is not clear whether the
sources we are detecting at $z>5$ are representative of the entire
collapsed mass in a galactic halo at these redshifts, or simply high
surface brightness regions of strong star formation within larger,
lower surface brightness systems.

\section{AGN contamination}

What fraction of sources selected by our criteria are AGN? A simple
expectation of currently favoured structure-formation scenarios is
that the first galaxies form in the most overdense halos in the early
universe. These are also most likely to be the place where the first
supermassive black holes form, given the known correlation between central
black hole mass and spheroid mass ({\it e.g.} Gebhardt et~al., 2000,
Ferrerese \& Merritt 2000).  Clearly, in order to confront models of
early structure formation with data requires knowledge of what fraction
of these sources contain active AGN.  All the objects in Lehnert \&
Bremer (2003) with spectroscopically-determined redshifts had narrow
emission-lines and  showed no evidence for the broad lines expected from
relatively unobscured AGN. However, it is possible that some may have
contained optically obscured AGN (as observed by, for example, Giacconi
et al. 2001). These would be detectable in sufficiently deep X-ray
observations sensitive to rest-frame hard X-ray emission from the sources.

The CDF-S field has a 1Msec {\it Chandra} exposure (Rosati
et~al. 2002) publically available. We cross-correlated our source-list
with the catalogue of Giacconi et~al. (2002). None of our V-band
dropouts were among the catalogued sources. Having determined the
(small) offset between the X-ray coordinate frame and that of the ACS
from this cross-correlation, we then determined the 0.5-5keV X-ray
fluxes in 2 arcsec radius apertures centred on the positions of our 44
dropout sources. None were detected at a 3-$\sigma$ level of $8\times
10^{-6}$ counts sec$^{-1}$ above background and only one at 2-$\sigma$
(and one is expected from Poissonian noise in the background). The
distribution of counts in the apertures was indistinguishable from
Poissonian.  A total of 286 background plus object counts is
consistent with the expected background level of 306
counts. Consequently, no individual source was detected, and the
collective mean flux of a dropout source is less than 2 counts (at
$3-\sigma$). These limits corresponds to a flux of $7\times 10^{-17}$
erg cm$^{-2}$s$^{-1}$ for an individual source, and $2\times 10^{-17}$
erg cm$^{-2}$s$^{-1}$ for the mean source. At $z\sim 5.3$ these give a
$>2.5$ keV luminosity of $<2 \times 10^{43}$ erg s$^{-1}$ and $<5
\times 10^{42}$ erg s$^{-1}$ for individual and mean sources
respectively. This rules out contamination by powerful AGN since AGN
are generally prominent X-ray sources. These results are consistent
with those of Barger et al. (2003) but also extend them by going over
a magnitude deeper in the optical selection of sources allowing us to
investigate the most heavily obscured sources which have been shown to
contribute substantially to the total X-ray emission (e.g., Giacconi
et al. 2001) and by investigating the X-ray emission from
significantly more high redshift sources. These results are entirely
consistent with the sources having comparable X-ray luminosities to
those of $z=3$ Lyman break galaxies, which have L$_x \sim 3\times
10^{41}$ erg s$^{-1}$ above 2.5 keV (Brandt et~al. 2001).  The most
X-ray luminous starburst galaxies have similar luminosities to this
(e.g., Moran, Lehnert \& Helfand 1999). Given that our sources are
selected to have the colours of unobscured starbursts at $z>5$, any
future X-ray detection at this low level can be attributed to the
starburst rather than a weak AGN. Such a detection would require a
$>10$Msec {\it Chandra} exposure.

All of the above reinforces one of the conclusions of Lehnert \& Bremer
(2003) (and of Barger et al. 2003 for brighter optical sources), {\it
viz.} that emission from AGN played little part in the final stages of
the reionization of the Universe. Only if there were an early dominant
population of AGN that had completely died out by $z=5-6$, when strong
star formation had been initiated, could AGN have played a major part
in the reionization of the Universe.


\section{Conclusions}

We have selected a sample of 44 objects from the ACS data of the CDF-S
with properties consistent with galaxies at $z>5$ by using photometric
cuts to include objects similar to those spectroscopically confirmed
to be at $z>5$ by Lehnert \& Bremer (2003). The sample is likely to
contain a small number of contaminating sources at lower redshift, and
certainly does not contain all sources at $z>5$ due to its $i-$band
flux limit. Specifically, it will miss most objects of the type
selected by Stanway et~al. as they had $i-z>1.5$ in order to detect
galaxies at $z\sim 6$ and such objects are rare at $i<26.3$.

We found that these sources were often resolved on scales of $\sim
0.2-0.3$ arcsec, with several appearing double or with extended lower
surface brightness regions close to them, quite similar to the $z=5.3$
source in the HDF-N described by by Spinrad et~al. (1998). The candidate
$z>5$ dropout galaxies appear to be smaller than those studied by
Lowenthal et~al. (1997) at $z=3$, likely indicating that the scale
lengths of galaxies decrease with increasing redshift.

The distribution of the selected objects over the 150 arcmin$^2$ area
is not uniform, there is a patch of area $\sim 25$ arcmin$^2$ where
there are no selected sources. Comparing nearest-neighbour
distributions of these sources and randomly selected samples of the
same number of faint, non-dropout sources leads to no detectable
differences. Similarly, ``holes'' of a similar size occasionally arise
in the areal distributions of the random samples. Nevertheless,
structures on this scale imply that fields considerably larger than
the ACS CDF-S field are required to probe the clustering and density
of high redshift sources with any degree of reliably. Interpretation
(and indeed field-placement) of smaller fields are complicated by such
a non-uniform distribution of sources.

None of the selected sources was conclusively detected as an X-ray
source, and the total flux from the 44 sources is also consistent with
zero. This limits the $>2.8$ keV X-ray luminosities of the individual
sources at $z\sim 5.3$ to considerably below $2\times 10^{43}$ erg
s$^{-1}$ (3-$\sigma$ upper limit on an individual source). This rules
out strong AGN contamination for the bulk of the sources. Given the short
timescale between $z\sim 5.3$ and the end of reionization at $z\sim 6$,
this supports the idea that AGN made little contribution to the final
stages of the reionization of the Universe.

\section{Acknowledgements}

We thank the GOODS teams at STScI and ESO for providing the public
reduced data. MNB, SP and IW acknowledge funding from the Leverhulme
Trust. We thank Mark Birkinshaw for helpful advice.


\begin{thebibliography}{}

\bibitem[\protect\citeauthoryear{Barger et al.}{2003}]{Barger03}Barger
A.~J., et al. 2003, ApJ, 584, L61

\bibitem[\protect\citeauthoryear{Bertin \& Arnouts}{1996}]{Bertin96}
Bertin E., Arnouts S., 1996, A\&AS, 117, 393

\bibitem[\protect\citeauthoryear{Bouwens et al.}{2003}]{Bouwens03}
Bouwens R.~J., et al., 2003, ApJ, in press (astro-ph/0306215)

\bibitem[\protect\citeauthoryear{Brandt et al.}{2001}]{Brandt01} Brandt
W.~N., Hornschemeier A.~E., Schneider D.~P., Alexander D.~M., Bauer F.~E.,
Garmire G.~P., Vignali C., 2001, ApJ, 558, L5

\bibitem[\protect\citeauthoryear{Bunker et al.}{2003}]{Bunker03} Bunker
A.~J., Stanway E.~R., Ellis R.~S. McMahon R.~G., McCarthy P.~J., 2003,
MNRAS, submitted (astro-ph/0302401)

\bibitem[\protect\citeauthoryear{Cimatti et al.}{2002}]{Cimatti02}
Cimatti A., et al., 2002, A\&A, 381, L68

\bibitem[\protect\citeauthoryear{Cuby et al.}{2003}]{Cuby03} Cuby J.-G.,
Le Fevre O., McCracken H., Cullandre J.-C., Magnier E., Meneux B.,  2003,
A\&A, in press (astro-ph/0303646)

\bibitem[\protect\citeauthoryear{Dickinson \&
Gialalisco}{2002}]{Dickinson02} Dickinson M., Giavalisco M., 2002,
in Bender R., Renzini A., eds, The Mass of Galaxies at Low and High
Redshift. Springer, Dordrecht, p.~324

\bibitem[\protect\citeauthoryear{Ferrarase \&
Merritt}{2000}]{Ferrarrese00} Ferrarese L., Merritt D., 2000, ApJ, 539, L9

\bibitem[Fioc \& Rocca-Volmerange\null (1997)]{Fioc97} Fioc M.,
Rocca-Volmerange B., 1997, A\&A, 326, 950

\bibitem[\protect\citeauthoryear{Ford et al.}{2003}]{Ford03} Ford H.~C.,
et al., 2003, Proc. SPIE, 4854, 81

\bibitem[\protect\citeauthoryear{Fruchter \& Hooke}{2002}]{Fruchter02}
Fruchter A., Hook R., 2002, PASP, 114, 144

\bibitem[\protect\citeauthoryear{Gebhardt et al.}{2000}]{Gebhardt00}
Gebhardt K., et al., 2000, ApJ, 539, L13

\bibitem[\protect\citeauthoryear{Giacconi et al.}{2001}]{Giacconi01}
Giacconi R., et al., 2001, ApJ, 551, 664

\bibitem[\protect\citeauthoryear{Giacconi et al.}{2002}]{Giacconi02}
Giacconi R., et al., 2002, ApJS, 139, 369

\bibitem[\protect\citeauthoryear{Iwata et al.}{2003}]{Iwata03} Iwata I.,
Ohta K., Tamura N., Ando M., Wada S., Watanabe C., Akiyama M., Aoki K.,
2003, PASJ, 55, 415

\bibitem[\protect\citeauthoryear{Kron}{1978}]{Kron78} Kron R.G., 1978,
PhD thesis, Univ.\ California at Berkeley

\bibitem[\protect\citeauthoryear{Lehnert and Bremer}{2003}]{Lehnert03}
Lehnert M.~D., Bremer M.~N., 2003, ApJ, in press (astro-ph/0212431)

\bibitem[\protect\citeauthoryear{Lowenthal et al.}{1997}]{Lowenthal97}
Lowenthal J.~D., et al., 1997, ApJ, 481, L673

\bibitem[\protect\citeauthoryear{Moran, Lehnert, \&
Helfand}{1999}]{Moran99} Moran,~E.~D., Lehnert,~M.~D., \&
Helfand,~D. 1999, ApJ, 526, 649

\bibitem[\protect\citeauthoryear{Roche et al.}{1998}]{Roche98} Roche N.,
Ratnatunga K., Griffiths R.~E., Im M., Naim A., 1998, MNRAS, 293, 157

\bibitem[\protect\citeauthoryear{Rosati et al.}{2002}]{Rosati02} Rosati
P., et al., 2002, ApJ, 566, 667

\bibitem[\protect\citeauthoryear{Spinrad et al}{1998}]{Spinrad98}
Spinrad H. et al., 1998, AJ, 116, 2617

\bibitem[\protect\citeauthoryear{Stanway et al.}{2003}]{Stanway03}
Stanway E.~R., Bunker A.~J., McMahon R.~G., 2003, MNRAS, 342, 439


\end{thebibliography}
\end{document}